\documentstyle[epsfig,twocolumn]{esapub}

\def\ga{\mathrel{\mathchoice {\vcenter{\offinterlineskip\halign{\hfil
   $\displaystyle##$\hfil\cr>\cr\sim\cr}}}
   {\vcenter{\offinterlineskip\halign{\hfil$\textstyle##$\hfil\cr>\cr\sim\cr}}}
   {\vcenter{\offinterlineskip\halign{\hfil$\scriptstyle##$\hfil\cr>\cr\sim\cr}}}
   {\vcenter{\offinterlineskip\halign{\hfil$\scriptscriptstyle##$\hfil\cr>\cr
   \sim\cr}}}}}

\newcommand{\pPm}{ { _{_{+}}\!\!\!\,\Pi\!\!_{_{ -}} } }
\newcommand{\dpPm}{ { _{_{+}}\!\!\!\,\Pi\!\!_{_{ -}\!*} } }
\newcommand{\mPp}{{_{_{ -}}\!\!\!\Pi\!_{_{ +}}}}
\newcommand{\dmPp}{ { _{_{ -}}\!\!\!\Pi\!_{_{ +}\!*} } }
\newcommand{\Dpm}{{_{_{ +}}\!\!{\cal D}\!\!\!_{_{ -}}}}
\newcommand{\Dmp}{{_{_{ -}}\!\!\!{\cal D}\!\!_{_{ +}}}}

\newcommand{\be}{\begin{equation}}
\newcommand{\ee}{\end{equation}}
\newcommand{\beq}{\begin{eqnarray}}
\newcommand{\eeq}{\end{eqnarray}}

\newcommand{\aeta}[3]{ 19#1,  A\&A  #2, \rm #3}
\newcommand{\aetap}[1]{ 19#1,  A\&A \rm in press}

\newcommand{\apj}[3]{ 19#1,  ApJ  #2, \rm #3}

\newcommand{\jgr}[3]{ 19#1,   JGR  #2, \rm #3}
\newcommand{\jgrp}[1]{ 19#1,   JGR \rm in press }
\newcommand{\pfl}[3]{ 19#1,   Phys. Fluids  #2, \rm #3}
\newcommand{\paspcs}[3]{ 19#1,   PASP conf. ser.  #2, \rm #3}

\newcommand{\solp}[3]{ 19#1,  Solar Phys.  #2, \rm #3}

 \newcommand{\bm}[1]{\mbox{\boldmath $#1$}}

\begin{document}

\setlength{\parindent}{0pt}
\setlength{\parskip}{ 10pt plus 1pt minus 1pt}
\setlength{\hoffset}{-1.5truecm}
\setlength{\textwidth}{ 17.1truecm }
\setlength{\columnsep}{1truecm }
\setlength{\columnseprule}{0pt}
\setlength{\headheight}{12pt}
\setlength{\headsep}{20pt}
\pagestyle{esapubheadings}

\title{\bf QUASI-SEPARATRIX LAYERS: REFINED THEORY AND\\ 
       ITS APPLICATION TO SOLAR FLARES}

\author{{\bf V.S. Titov$^1$, P.~D\'emoulin$^2$ \ and G.~Hornig $^{1}$} 
\vspace{2mm} \\
$^1$Theoretische Physik IV, Ruhr-Universit\"{a}t Bochum,
              44780 Bochum, Germany \\
$^2$DASOP, Observatoire de Paris-Meudon, F 92195 Meudon cedex, France  }

\maketitle

\begin{abstract}

 Although the analysis of observational data indicates  that
quasi-separatrix layers (QSLs)  have to play an important role
in magnetic configurations  of solar flares, the corresponding
theory is only at an initial stage so far.  In particular,  there is still a need in
a proper definition of QSL.  This problem is analyzed here on the basis
of geometrical properties of the mapping produced by the field lines which
connect photospheric areas of positive and negative magnetic polarities of active
regions.
  In general, one can find on the photosphere a unique pair of
locally perpendicular line elements at one footpoint of a given field line, so
that this pair is mapped into a similar pair of elements at the other footpoint.
  Along the directions of these elements the field line mapping only stretches and
compresses  the corresponding displacements of the neighboring field-line
footpoints.
 This fact enables us to get a remarkably clear and concise definition of QSL
as a volume filled by coronal field lines for which the ratio of the
corresponding  stretching and compressing coefficients is anomalously large.
 The new definition is also compared with the ones previously introduced
by other authors.

The theory is applied to flare events of the so-called sigmoid-type, using
an analytical model of a twisted force-free configuration
(\cite{TitDem99a}, b).   It is shown that such a configuration
may contain a QSL even if genuine separatrix surfaces are absent.  
 Identifying QSLs is an essential prerequisite for understanding
the mechanism of  magnetic energy release in this type of flares.
 It is also demonstrated that the magnetic field under study has a Hamiltonian
structure, which makes it possible to reveal a geometrical reason
for the appearance of  QSL in this case.
 \vspace {5pt} \\

  Key~words: Quasi-separatrix layers; magnetic topology; solar flares.

\end{abstract}

\section{INTRODUCTION}


Investigations of coronal magnetic  fields extrapolated from
photospheric magnetograms do not always show a relation of solar
flares with the separatrices caused by magnetic null points
(\cite{Dem94}). 
  There is, however, a more systematic spatial correlation between the 
locations of energy release in flares and the regions of strong
variation of the field line connectivity (e.g. \cite{Man95}, \cite{Dem97}). 
  Such regions, called quasi-separatrix layers (QSLs), are thought to be 
the plausible places for the magnetic reconnection process (\cite{PriDem95}).   


 In most of the coronal volume  the quasi-static conditions are
fulfilled, so that the magnetic field evolves through a sequence of
force-free  equilibria.
 These conditions, however, may easily break down in QSLs,  where due to a
strong variation of the field line connectivity the rearrangement of
the field lines during the evolution of the configuration may occur faster
than in other places.
  This in turn implies  a locally large acceleration of plasma and hence
a locally unbalanced and enhanced Lorentz force with
corresponding concentrations of the current density.
  So ultimately the inertia of plasma may cause the formation of current layers in QSLs.
  The importance of inertia in the current layers at the QSLs is also
followed from exact solutions of linearized MHD equations describing a
quasi-static evolution of inhomogeneous magnetic fields (\cite{InvTit97}).


 The paper is organized as follows.  In Sec.~2, we discuss the
difference between separatrix surfaces and QSLs together with the
primary definition of QSLs.   In Sec.~3, the local geometrical
properties of the magnetic connectivity are studied.  The new
geometrical measure for QSLs is derived and compared with the old
ones.  Sec.~4 contains an application of this theory to a model of flaring
twisted configurations possessing a Hamiltonian
structure, which plays an important role in the formation of QSLs.
The conclusions are summerized in Sec.~5.

\section{THE QSL AND ITS PRIMARY DEFINITION}


  The magnetic field lines in solar active regions normally connect
domains of positive and negative polarity of the photospheric plane, say,~$z=0$.
  Let the location of their footpoints in this plane be  represented
depending on the polarity by the radius-vector ${\bf r}_{+}=(x_{+}, y_{+})$ or 
${\bf r}_{-} = (x_{-}, y_{-})$.
  The connections of the footpoints by the field
lines determine two mutually inverse mappings  $\pPm:\, {\bf
r}_{+}  \mapsto {\bf r}_{-}$  and $\mPp:\,{\bf r}_{-}
\mapsto  {\bf r}_{+}$.
  We shall simply use $\Pi$  if we refer to aspects valid for both mappings.
  Also the functional forms $  (X_{-}({\bf r}_{+}), Y_{-}({\bf
r}_{+}))\equiv {\pPm}({\bf r}_{+}) $ and $  (X_{+}({\bf r}_{-}), Y_{+}({\bf
r}_{-}))\equiv {\mPp}({\bf r}_{-}) $ will  be used further for the mappings.


  The mapping  $\Pi$ is discontinuous at the footpoints of the field lines
threading magnetic nulls in the corona or touching the photosphere,
since the magnetic flux tubes enclosing such field lines are splitted
at the nulls or at the touching points (\cite{See86}).
 The corresponding discontinuities serve as indicators for
the separatrix field lines and surfaces.
 It is worth to emphasize that the coordinates  $(x_{\pm},  y_{\pm})$
in this case need not to be Cartesian because the discontinuities are
revealed in any system of coordinates irrespective of the metrics.


  However, with the help of the metrics or Cartesian coordinates one
can determine not only the genuine separatrices but also the QSLs.
  The integrity of the flux tubes is preserved within the QSLs
and so the mapping  $\Pi$ remains continuous
at the corresponding footpoints, but the shape of their cross-sections
strongly change along the flux tubes.
  Thus, instead of the true discontinuities in $\Pi$ at the
intersection of the genuine separatrices with the photosphere,
there are continuous but rapid variations in $\Pi$ at the
photospheric cross-sections of QSLs.
  These variations can only be detected by using the metrics, which
enables us to measure and compare the distances between the footpoints 
in one polarity and corresponding footpoints in the other polarity.
  In this respect QSLs and separatrices are qualitatively
different objects.
  Indeed, ignoring the above metrical information about $\Pi$ and
using a proper continuous change of coordinates, it is possible to
eliminate  the rapid variations in $\Pi$ and thereby the QSLs
themselves, while the discontinuities of $\Pi$ and hence the
corresponding separatrices are not removable in this way.
  However, for physical processes with characteristic length
scales much less than the QSL thickness such a QSL must be as important
as a genuine separatrix.


 For the determination of the QSLs \cite*{PriDem95} proposed to
use the function (called ``the norm'')  $N({\bf r}_{+})$ or $N({\bf
r}_{-})$,  which in Cartesian coordinates are
 \begin{eqnarray}
  N({\bf r}_{\pm})   =\left[ \left( {\partial X_{\mp} \over \partial
x_{\pm}} \right)^{2} + \left( {\partial X_{\mp} \over \partial y_{\pm}}
\right)^{2} +\right.\qquad \qquad && \nonumber \\
  \qquad \qquad \left.\  \left( {\partial Y_{\mp} \over \partial x_{\pm}}
\right)^{2} + \left( {\partial Y_{\mp} \over \partial y_{\pm}}
\right)^{2} \right]^{1/2}  \equiv N_{\pm} \, .&&
     \label{N1}  
 \end{eqnarray}
 It was suggested that $N({\bf r}_{\pm}) \gg 1$ at the footpoints of the 
field lines belonging to QSLs and $N({\bf r}_{\pm}) \approx 1$ otherwise.
  Yet this norm in application to different footpoints of the same
field line  yields generally different values  $N_{+}$ and  $N_{-}$,  which
leads to an ambiguity in the determination of QSLs.
   This disadvantage of the norm indicates that the adequate measure for 
QSLs must be invariant to the choice of the mapping  $\pPm$ or $\mPp$.
  Below we find such a measure by analyzing geometrical properties of
the field line connectivity.

\section{REVISED DEFINITION OF QSL}

  The mapping  $\pPm$  or $\mPp$ is locally
described by its differential  $\dpPm$ or $\dmPp$, respectively,
which is a linear mapping from the plane tangent to
the  photosphere at one footpoint to a similar plane at the other footpoint.
 These differentials are represented by the corresponding, mutually
inverse,  Jacobian matrices
 \begin{eqnarray}
 \Dpm
  = \left( \begin{array}{cc}
 {\partial X_{-}\over \partial x_{+}}
    & {\partial X_{-}\over \partial y_{+}}\\
 {\partial Y_{-}\over \partial x_{+}}
    & {\partial Y_{-}\over \partial y_{+}}
           \end{array} 
    \right)
 \equiv \left( \begin{array}{cc}
    a    & b\\
    c    & d
           \end{array}
    \right)
  \label{Dpm}
 \end{eqnarray}
and
 \begin{eqnarray}
 &&\hspace{-5mm}\Dmp
  = \left( \begin{array}{cc}
 {\partial X_{+}\over \partial x_{-}}
    & {\partial X_{+}\over \partial y_{-}}\\
 {\partial Y_{+}\over \partial x_{-}}
    & {\partial Y_{+}\over \partial y_{-}}
           \end{array} 
    \right)
 =\Delta^{-1}_{+}\left( \begin{array}{cc}
    d    &-b\\
    -c    & a
           \end{array}
    \right)\!\!  ,  \ 
  \label{Dmp}  \\
 &&\hspace{-5mm}\Delta_{+} = a d- b c \equiv \det \Dpm.
	\label{det}
 \end{eqnarray}
 We assume hereafter that  $(x_{\pm}, y_{\pm})$ are
measured in one Cartesian system of coordinates covering the whole
photospheric plane.
  Eqs. (\ref{Dpm}) and (\ref{Dmp}) show that it is sufficient to have
only one of these matrices for a local description of the magnetic connectivity.

\begin{figure*}[!ht]
  \begin{center}
    \leavevmode
  \centerline{\epsfig{file=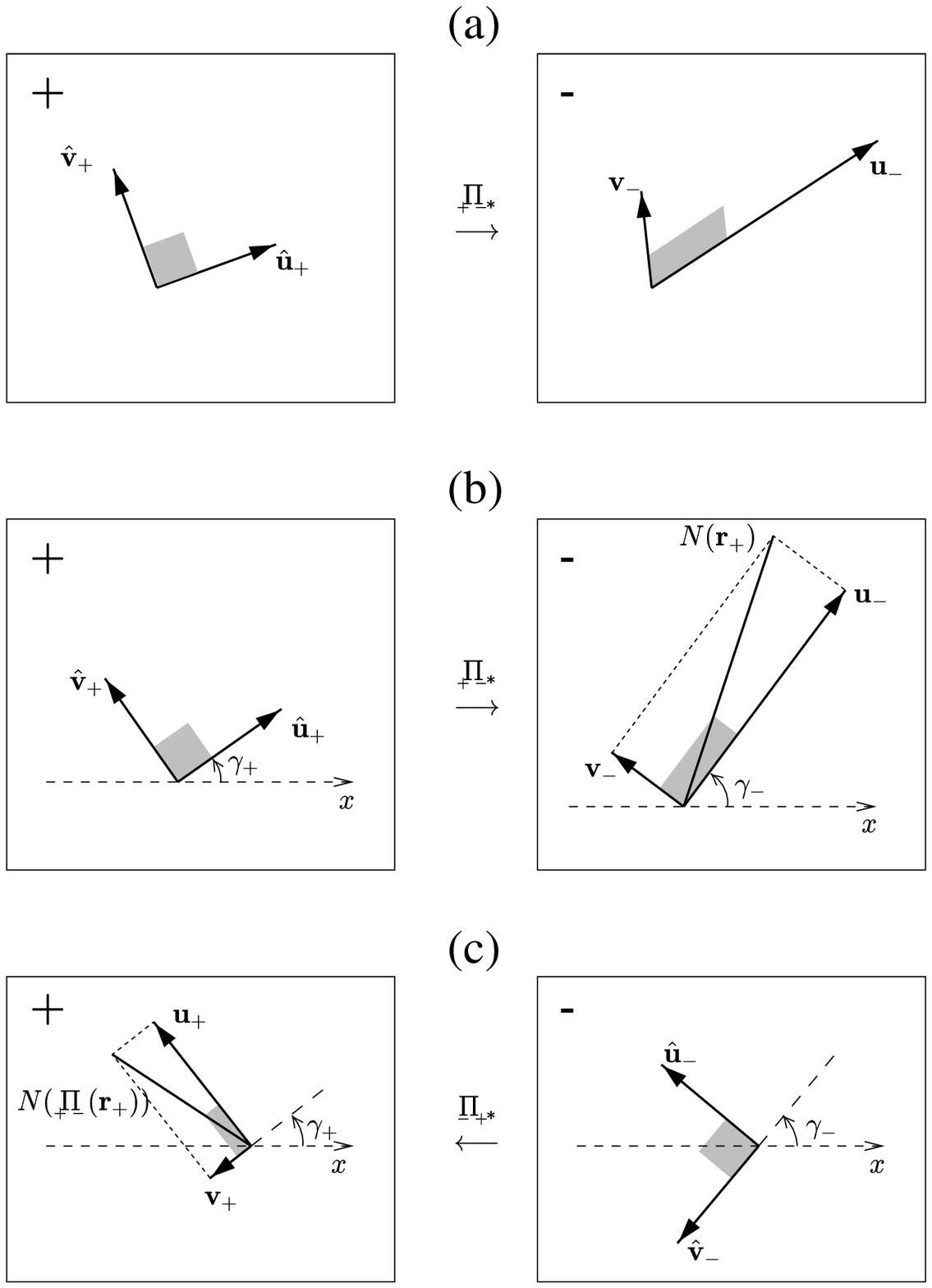,width=17cm}}
  \end{center}
  \caption{\em The mapping of an orthonormal basis  $(\hat{{\bf
u}}_{+}, \hat{{\bf v}}_{+})$  by the differential  $\dpPm$
(a,b): its image basis  $({\bf u}_{-}, {\bf v}_{-})$ is in general
non-orthonormal (a), but it is orthogonal for a special orientation of 
$(\hat{{\bf u}}_{+}, \hat{{\bf v}}_{+})$ (b) when  $|{\bf u}_{-} +
{\bf v}_{-}|$ determines the norm  $N({\bf r}_{+})$.
  The inverse differential  $\dmPp$ maps the orthonormal basis
$(\hat{{\bf u}}_{-},  \hat{{\bf v}}_{-})$, rotated with respect to
$({\bf u}_{-}, {\bf v}_{-})$ on  $\pi/2$, into the orthogonal basis
$({\bf u}_{+}, {\bf v}_{+})$ (c), so that  $|{\bf u}_{+}|/|{\bf v}_{+}|
= |{\bf u}_{-}|/|{\bf v}_{-}|$ but  $|{\bf u}_{+} + {\bf v}_{+}|= 
N(\pPm({\bf r}_{+})) \not = N({\bf r}_{+}) $.  The shaded areas show
how these properties of the mappings manifest in the corresponding
cross-sections of thin flux tubes.}
  \label{f:tp}
\end{figure*}

  An orthonormal vector basis  $(\hat{\bf u}_{+}, \hat{\bf v}_{+})$ in 
the tangent plane at some footpoint  ${\bf r}_{+}$ is generally mapped 
by  $\dpPm$ into a non-orthonormal basis  $({\bf
u}_{-}, {\bf v}_{-})$ with   ${\bf u}_{-} = \Dpm \hat{\bf
u}_{+}$ and ${\bf v}_{-} = \Dpm\hat{\bf v}_{+}$ in the
corresponding tangent plane at  $\pPm({\bf r}_{+})$
(Figure \ref{f:tp}a). 
  One can choose, however, a special orientation of
$(\hat{\bf u}_{+}, \hat{\bf v}_{+})$ such that the basis $({\bf
u}_{-}, {\bf v}_{-})$ becomes orthogonal (Figure \ref{f:tp}b).
   Indeed, let the angles  $\gamma_{+}$ and  $\gamma_{-}$,
respectively,  determine the directions of  $\hat{\bf u}_{+}$ and  ${\bf u}_{-}$
with respect to the  $x$-axis of the global Cartesian system of
coordinates.
   ${\cal R}_{\gamma_{+}}$ and  ${\cal R}_{\gamma_{-}}$
are the corresponding matrices of rotations.
  It follows then from Figure \ref{f:tp}b  that  $|{\bf u}_{-}|\, {\cal
R}^{-1}_{\gamma_{+}}\, \hat{\bf u}_{+} = {\cal R}^{-1}_{\gamma_{-}}\,
\Dpm\,\hat{\bf u}_{+}$  and $|{\bf v}_{-}|\, {\cal
R}^{-1}_{\gamma_{+}}\, \hat{\bf v}_{+} = {\cal R}^{-1}_{\gamma_{-}}\,
\Dpm\, \hat{\bf v}_{+}$ or, in other words,
 \begin{eqnarray}
  {\rm diag}(|{\bf u}_{-}|, |{\bf v}_{-}|) = {\cal
R}^{-1}_{\gamma_{-}} \, \Dpm \, {\cal R}_{\gamma_{+}}.
	\label{diag}
 \end{eqnarray}
  It means that the two non-diagonal elements at the right hand side
of (\ref{diag})  must vanish.   
 This condition can be satisfied by
 \begin{eqnarray}
  \gamma_{+} -\gamma_{-} &=& \arg\xi,
	\label{gm} \\
 \gamma_{+} +\gamma_{-} &=& \arg\zeta,
	\label{gp} 
 \end{eqnarray}
 where the function  $\arg$ determines the
arguments  of the complex values (${\rm i}=\sqrt{-1}$)
 \begin{eqnarray}
  \xi& =& a+d+{\rm i}\, (b-c), 
	\label{zm} \\
  \zeta &= &a-d+{\rm i}\, (b+c).
	\label{zp}
 \end{eqnarray}
  This solution for $\gamma_{+}$ and  $\gamma_{-}$  can be checked by 
substitution of  (\ref{Dpm}) and  the corresponding expressions for the rotation
matrices in (\ref{diag}).
  Other solutions are also possible but not of interest, since they are simply the result of
mirror reflections about the directions given by (\ref{gm}) and
(\ref{gp}).

Eqs. (\ref{diag})--(\ref{zp}) and (\ref{Dpm}) yield
 \begin{eqnarray}
   |{\bf u}_{-}|& = &  ( |\xi| + |\zeta|)/2,
	\label{mum} \\
   |{\bf v}_{-}| & = &  ( |\xi| - |\zeta|)/2,
	\label{mvm}
 \end{eqnarray}
which means that  $|{\bf u}_{-}|\ge |{\bf v}_{-}|$ in the solution
under consideration.
  By using these expressions and (\ref{N1})--(\ref{Dpm}) one can obtain
 \begin{eqnarray}
     |{\bf u}_{-} +{\bf v}_{-}|= (a^{2} + b^{2} + c^{2} +d^{2})^{1/2} \equiv N_{+} .
	\label{N2}
 \end{eqnarray}
  Thus the norm $N_{+}$ determines simply the length of the rectangular
diagonal constructed on the vectors  ${\bf u}_{-}$  and  ${\bf v}_{-}$
(Figure  \ref{f:tp}b).

It is worth to determine similar characteristics for the
reverse differential  $\dmPp$ at ${\bf r}_{-} = \pPm({\bf
r}_{+})$ .
  The simplest way to do this is just to change in
(\ref{gm})--(\ref{mvm}) the superscripts $\pm$ on  $\mp$ and the elements
of $\Dpm$ (\ref{Dpm}) to the corresponding elements of  $\Dmp$
(\ref{Dmp}).
  Then we obtain for the complex values similar to (\ref{zm})  and (\ref{zp}) 
 \begin{eqnarray}
  \tilde{\xi} &=& [d+a-{\rm i}\, (b-c)]/\Delta_{+} \equiv
\bar{\xi}/\Delta_{+},  
	\label{tx} \\
 \tilde{\zeta} &=& [d-a-{\rm i}\, (b+c)]/\Delta_{+} \equiv -\zeta/\Delta_{+},
	\label{tz}
 \end{eqnarray}
 yielding the following equations for the angles:
 \begin{eqnarray}
  \tilde{\gamma}_{-} -\tilde{\gamma}_{+} &=& \arg\bar{\xi} \equiv -\arg \xi,
	\label{tgm} \\
 \tilde{\gamma}_{-} +\tilde{\gamma}_{+} &=& \arg\zeta + \pi.
	\label{tgp}  
 \end{eqnarray}
  They show that the orthonormal basis  $(\hat{\bf u}_{-}, \hat{\bf
v}_{-})$ corresponding to this solution is turned on  $\pi/2$ with
respect to the basis  $({\bf u}_{-}, {\bf v}_{-})$ and the same is
valid for $({\bf u}_{+}, {\bf v}_{+})$  and $(\hat{\bf u}_{+}, \hat{\bf
v}_{+})$ (Figure \ref{f:tp}b, c).
  In application to (\ref{mum}) and (\ref{mvm}) the considered
transformation gives
\begin{eqnarray}
   |{\bf u}_{+}|& = & 1/|{\bf v}_{-}|,
	\label{mup} \\
   |{\bf v}_{+}| & = &  1/|{\bf u}_{-}|.
	\label{mvp}
 \end{eqnarray}
  So the norm in the negative polarity at the footpoint  ${\bf r}_{-}
=  \pPm({\bf r}_{+})$ is
 \begin{eqnarray}
  N_{-} \circ \pPm =N_{+} /|\Delta_{+}|.
	\label{N3}
 \end{eqnarray}
 This consideration shows that the determination of QSLs by
means of the norms  $N_{+}$ and  $N_{-}$ must really lead to different 
results if  $|\Delta_{+}| \not = 1$.
  Combining of these results in order to find the ``proper'' QSL
is in fact equivalent to finding only in the positive polarity the regions where 
the functions  $N_{+}$ and  $N_{+}/|\Delta_{+}|$ acquire anomalously large
values.
  Yet the simultaneous using of two different functions for
characterizing QSLs does not look as a theoretically well-founded
approach.

 One can resolve this difficulty by noticing, first, that the mapping 
$\Pi$ can locally be described by  $|{\bf u}_{-}|$, $|{\bf v}_{-}|$,
$\gamma_{+}$ and $\gamma_{-}$, where only $|{\bf u}_{-}|$ and $|{\bf
v}_{-}|$ determine the value of footpoint displacements, while
$\gamma_{+}$ and $\gamma_{-}$ define their directions.
  So it would be natural if the required characteristics is a function of $|{\bf
u}_{-}|$ and $|{\bf v}_{-}|$ only.
  We think that the ratio $|{\bf u}_{-}|/|{\bf v}_{-}|\ge 1$ is a more
suitable quantity for describing QSLs than the considered above norm.
  According to (\ref{mup}) and (\ref{mvp}) (see also Figure \ref{f:tp}b,
c) it coincides with the ratio  $|{\bf u}_{+}|/|{\bf v}_{+}|$ and
thereby characterizes the magnetic connectivity itself rather than one 
of the mappings  $\pPm$ or  $\mPp$.
  It determines the degree of flattening of the elementary flux tubes
at their photospheric ends (see shaded regions in Figure \ref{f:tp}b,c), so the tubes
characterized by extremely high values 
of this quantity should be referred to as QSL.
  By using (\ref{zm})--(\ref{mvm}) one can derive that
 \begin{eqnarray}
  |{\bf u}_{-}| /|{\bf v}_{-}| &=& Q/2 + \sqrt{Q^2/4-1} ,
	\label{ar}  \\
   Q&=&N^{2}_{+}/|\Delta_{+}|, 
	\label{q}
 \end{eqnarray}
which shows that $ |{\bf u}_{-}| /| {\bf v}_{-}| \approx Q $ 
for  $Q\gg 1$.
  It is seen from (\ref{N3}) and (\ref{q}) that
 $Q$ is simply a product of the norms $N_{+}$ and $N_{-}$ calculated at
the opposite ends of the field lines.
  Geometrically, the value  $Q$ is the ratio of the square of quadrat with the side 
 $N_{+}$  to the square of the rectangular based on the vectors  ${\bf 
u}_{-}$ and  ${\bf v}_{-}$, because  $|\Delta_{+}| = |{\bf u}_{-}|\cdot |{\bf 
v}_{-}|$.
  Note also that $Q \ge 2$, since $N_{+}$ is the length of the diagonal in this
rectangular.
 The expression for  $Q$ has an elegant form, invariant to the change  $+
\leftrightarrow -$ and simpler than (\ref{ar}), so it is reasonable to
define:
\centerline{
\parbox[][10mm][c]{80mm}{
  {\em The QSL is a flux tube consisting of magnetic field lines
with $Q \gg 1$\/}.} }
  One can see from (\ref{q}) that if the determinant  $\Delta_{+}$
varies in the photospheric plane as strong as  $N_{+}$, the QSLs
defined with the help of  $Q$ or  $N_{+}$ have to be different,
otherwise they must  be approximately the same.
   
  Some comments should also be added on another characteristics for
determining QSLs, namely, on the so-called differential flux volume
 \begin{eqnarray*}
  V({\bf r}_{+})= \int^{\pPm({\bf r}_{+})}_{{\bf r}_{+}}
{{\rm d}l  \over B} \equiv V(\pPm({\bf r}_{+})),
 \end{eqnarray*}
in which the integration is carried out along the corresponding field line.
  This value has appeared in the analysis of current sheet formations
 along separatrix surfaces in quasi-static evolutions of
$2{1  \over 2}$D magnetic configurations
(\cite{ZSB85},\cite{LW88}, \cite{VPA91}, \cite{VP92}).
  It has also been used (under the name ``delay function'') for studying
3D magnetic topology caused by the presence of null points (\cite{Lau93}).
  Recently \cite*{SB99} have shown that strong spatial variations of 
$V$ may cause a similar formation of current layers in
3D magnetic fields.
  So it seems natural at first sight to use this value for characterizing QSLs.
  However, if one fixes the footpoints of the magnetic field lines in a given
configuration and exposes the coronal volume to a smooth deformation, 
then $V$ will be changed depending on such a deformation, 
while the field line connectivity will remain the same.
  Thus, the quantity  $Q$ as a measure of the field line connectivity is better
than $V$, although this does not reduce the value of  $V$ in
understanding of the above mentioned current sheet formation.

 \section{HAMILTONIAN STRUCTURE AND QSL 
   IN FLARING TWISTED CONFIGURATIONS}

\begin{figure}[!ht] 
   \begin{center}
    \leavevmode
  \centerline{\epsfig{file=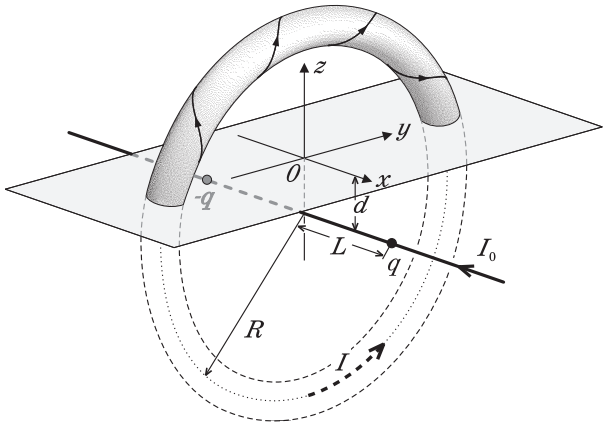,width=8cm}}
  \end{center}
   \caption{\em The scheme of the model for the twisted magnetic
configuration }
    \label{schm}
 \end{figure}

 The topology of a twisted magnetic configuration provides important
clues for understanding sigmoid-like flares (\cite{TitDem99a},~b).
  Such a configuration can be modeled by a force-free circular flux
tube with total current $I$, which is embedded into potential field
produced by a pair of magnetic charges $q$, $-q$ and a line current
$I_0$ (Figure \ref{schm}). 
  The charges are on a distance $2L$ from each other and lie together with
the current $I_0$ on the axis of symmetry of the configuration, which is
on the depth $d$ below the photospheric plane $z=0$. 
  Below this plane the sources and their fields have no a real
physical meaning: they are used only to construct the proper magnetic
field in the corona. 
  The photospheric vertical component of this field and current density of
the flux tube can be considered a posteriori as the corresponding boundary
conditions for the resulting configuration.

   The construction of this magnetic field excludes the existence of any null
point in our model, so the topology of the coronal field may be
non-trivial only due to the presence of the so-called  bald
patches  (BPs, see \cite{TPD93}), which are the segments of the
photospheric inversion line (IL) where the coronal field lines touch
the photosphere (\cite{See86}). 
 Thus, the separatrices, if present,  consist of only the field lines starting at the BPs. 
  The full topological analysis of the field then reduces to
determining the BPs and their associated separatrices. 
  Such an analysis has been made by \cite*{TitDem99b} for the scenario
in which the major radius $R$ of the flux tube (Figure \ref{schm})
grows,  starting from $R=d$, while the intensity and
positions of sub-photospheric sources remain constant.
   It was found that the BP appears first at some 
$R=R_{\rm a} > d$ and then grows in size with increasing~$R$.
   After reaching some $R=R_{\rm b} > R_{\rm a}$ the BP bifurcates at
the center of the configuration to give birth to a generalized
separator field line at the intersection of two separatrix surfaces.
   The gap between bifurcated parts of BP grows with $R$,  while these
parts shrink and eventually disappear at some $R=R_{\rm d}$ ($> R_{\rm
b}$),  where a transition to a topologically simple arcade-like configuration occurs.

   In spite of this simplicity, the resulting magnetic field might have
QSLs, which are  as important as the genuine separatrices under the
appropriate conditions.
  The aim of the further consideration is to check this possibility by 
using the above theory of QSLs for the same scenario of the magnetic field 
evolution.

 Here we use the cylindrical system of coordinates $(r_{\perp}, \theta, x)$,
whose axis is directed along $I_{0}$ and where $r_{\perp} \equiv |{\bf
r}_{\perp}|$ with ${\bf r}_{\perp} \equiv (0, y, z+d)$, $\theta$ is the
angle between  ${\bf r}_{\perp}$ and $\skew3\hat{\bf y}$, and
$\skew3\hat{\bm{\theta}}$ is the corresponding unit vector.
  Due to the axial symmetry about the line current $I_{0}$ (Figure \ref{schm})
the magnetic field can be written as
 \begin{eqnarray}
  {\bf B} = \nabla\times\left( \frac{\Psi}{r_{\perp}}
\skew3\hat{\bm{\theta}} \right) + B_{\theta}\, \skew3\hat{\bm{\theta}}\, ,
        \label{Bax}
 \end{eqnarray}
which, in particular, gives
 \begin{eqnarray}
  B_{x}={1 \over r_{\perp}} {\partial \Psi \over \partial r_{\perp} }
\, .
   \label{Psi}
 \end{eqnarray}
   It is convenient in our case to use $\theta$ for parameterizing
field lines, so that their equation ${\rm d}{\bf r}\times {\bf
B} = 0$ with  ${\rm d}{\bf r} = {\rm d}r_{\perp}\, \skew3\hat{\bf
r}_{\perp} + r_{\perp}\, {\rm d}\theta\, \skew3\hat{\bm{\theta}} + {\rm
d}x\, \skew3\hat{\bf x} $ yield
 \begin{eqnarray}
  {{\rm d}x\over{\rm d} \theta} & = & \frac{1}{B_{\theta}}\, {\partial \Psi\over
\partial r_{\perp}}\, ,
  \label{fle1} \\
    {{\rm d} r_{\perp} \over{\rm d} \theta} & = & -\frac{1}{B_{\theta}}\,
{\partial \Psi\over \partial x}\, .
  \label{fle2}
 \end{eqnarray}
These are the components of the field line equation along
$\skew3\hat{\bf  r}_{\perp}$ and $\skew3\hat{\bf x}$, respectively,
 while the component along $\skew3\hat{\bm{\theta}}$ requires ${\rm d}
\Psi=0$  along the field lines, which is simply a consequence of
(\ref{fle1})--(\ref{fle2}).

   The axial symmetry allows also to write the $x$-component of
the current density  ${\bf j}$ as
 \begin{eqnarray*}
  j_{x} = {1\over\mu_{0}  r_{\perp}}\, {\partial \over \partial r_{\perp}}
(r_{\perp} B_{\theta})\, ,
 \end{eqnarray*}
 from where after introducing the whole current $J = \int_{0}^{r_{\perp}} j_{x}\, 
2\pi r_{\perp}\, {\rm d}r_{\perp} $ through the perpendicular to the
$x$-axis disc of radius $r_{\perp}$ one can obtain
 \begin{eqnarray}
  B_{\theta} = {\mu_{0} J(\Psi)\over 2\pi r_{\perp}}\, .
        \label{Btg}
 \end{eqnarray}
   Here we put $J=J(\Psi)$, since the current lines lie on
magnetic surfaces $\Psi={\rm const}$ (note that in our model  ${\bf j}\; 
\| \; {\bf B}$ and  ${\bf B} \cdot \nabla \Psi = 0$ according to
Eq. (\ref{Bax})). 
 Moreover, this is also valid in more general case of magnetostatic configurations.
  By using (\ref{Btg}) and normalizing the
coordinates as $\skew3\tilde{r}_{\perp} = r_{\perp}/L$,
$\skew3\tilde{x} = x/L$ with subsequent omitting of the tilde, one can transform
(\ref{fle1})--(\ref{fle2}) to the following one-dimensional and
autonomous Hamiltonian system:
 \begin{eqnarray}
  {{\rm d}x\over{\rm d} \theta} & = &  {\partial  H \over \partial
\varpi}\, ,
  \label{Heq1} \\
    {{\rm d} \varpi \over{\rm d} \theta} & = & -{\partial  H\over
\partial x}\, ,
  \label{Heq2}
 \end{eqnarray}
 where $\theta$ is a time-like variable and instead of $r_{\perp}$ the new
momentum-like variable
 \begin{eqnarray}
  \varpi=\ln r_{\perp}
     \label{vpi}
 \end{eqnarray}
 is introduced together with the Hamiltonian function
  \begin{eqnarray}
    H = {2\pi \over \mu_{0} L}\, \int {{\rm d}\Psi \over J(\Psi)}\, .
     \label{Ham}
  \end{eqnarray} 
 The Hamiltonian structure is  revealed here only with the help of the
axial symmetry and the magnetostatic condition $J=J(\Psi)$, so it must
be inherent to both force-free and magnetostatic fields having the
axial symmetry.
  The analogous fact has been early established by \cite*{Lew90} for toroidal
geometry but we derived it in a much simpler way by using a different
Hamiltonian and canonical variables.

   This structure provides a powerful tool for investigating the
geometry of magnetic field lines in our model.
   Each field line can be considered now as a trajectory in the
extended phase space $(x,\varpi,\theta)$ of the autonomous one-dimensional
dynamical system  (\ref{Heq1}), (\ref{Heq2}).
  In the following it turns out that the most interesting behavior of the
field lines is outside of the flux tube, where the magnetic field is
purely potential and $J(\Psi) \equiv I_{0} $, so that
 \begin{eqnarray}
  B_{\theta} = {\mu_{0} I_{0} \over 2\pi r_{\perp}}\, .
        \label{Btp}
 \end{eqnarray}
  The Hamiltonian (\ref{Ham}) then reduces to
 \begin{eqnarray}
   H = {2\pi \Psi\over \mu_{0} I_{0} L}\, ,
        \label{Hp}
 \end{eqnarray}
 which is simply the poloidal flux function multiplied by a constant.
  This function is the following linear superposition:
  \begin{eqnarray}
   \Psi = \Psi_{q} + \Psi_{-q} + \Psi_{I},
        \label{Psp}
  \end{eqnarray}
 where the flux functions $\Psi_{\pm q}$ and $\Psi_{I}$, respectively,
correspond to the charges $\pm q$ and flux tube.
 The  $x$-component of the field produced by one charge is
   \begin{eqnarray}
    B_{\pm qx} = {\pm q(x \mp 1)\over [r_{\perp}^{2} + (x \mp
1)^{2}]^{3/2}}\, ,
        \label{Bqx}
   \end{eqnarray}
  which together with  (\ref{Psi})  enables us to obtain
  \begin{eqnarray}
   \Psi_{\pm q} = {\mp q\, (x\mp 1) \over
\sqrt{r_{\perp}^{2} + (x\mp 1)^{2}}  } \, .
        \label{Psq}
  \end{eqnarray}
   Let the minor radius  $a$ of the flux tube be much less than its
major radius  $R$, then  $\Psi_{I}$ must be well approximated outside
the tube by the flux function of the circle coil of radius $R$ with current~$I$.
  Due to the axial symmetry $\Psi_{I}(r_{\perp},x) = r_{\perp}\,
A_{I}(r_{\perp},x)$, where $A_{I}$ is one non-vanishing $\theta$-component of
the vector-potential
 \begin{eqnarray*}
  {\bf A}_{I}({\bf r}) = {\mu_{0} I\over 4\pi} \oint {{\rm d}{\bf
r}^{\prime}\over |{\bf r} - {\bf r}^{\prime}  |}\, ,
 \end{eqnarray*}
  in which the integration is made over the whole coil.
 After some transformations it gives
 \begin{eqnarray}
  \Psi_{I} = {\mu_{0} I \over \sqrt{8}} (LRr_{\perp}\upsilon)^{1/2}
\Phi(\upsilon),
        \label{PsI}
 \end{eqnarray}
 where
 \begin{eqnarray}
   \upsilon & = & {2 R/L\, r_{\perp} \over R^{2}/L^{2} + r_{\perp}^{2}
     +x^{2}}\, ,
        \label{ups} \\
  \Phi(\upsilon) & = & \frac{1}{\pi} \int\limits_{0}^{\pi} {\cos\theta\, {\rm
     d}\theta \over (1 - \upsilon \cos\theta)^{1/2}}\, .
        \label{Phi}
 \end{eqnarray}
   One can verify that $\Phi(\upsilon)$  is approximated by
 \begin{eqnarray}
  \Phi(\upsilon) \approx 0.043\, \upsilon - 0.089\, \upsilon^{2} - 0.207\,
\ln(1-\upsilon) 
        \label{Pha}
 \end{eqnarray}
  with the relative error $\le 1\%$ on the interval $[0, 0.9]$, which
is good enough for our further purposes.

  The Lorenz force $F_q  =  -{ 2 q L I
(R^2+L^2)^{-3/2} }$ caused by the interaction of the current $I$ with
the fields ${B}_{\pm qx}$ and the self-force ${F}_I = \mu_0 I^2 ( \ln
{8R / a} -5/4 ) / (4\pi R ) $ of the flux tube due to its curvature
counterbalance one another if the current is
 \begin{eqnarray}
    I = { 8 \pi q L R (R^2+L^2)^{-3/2}
   \over \mu_0 \left[ \ln (8R/a)  -5/4  \right] }
 \label{Ieq} \,.
 \end{eqnarray}
 The expression for  $F_{I}$ used here corresponds to the simplest
case when the equilibrium current $I$  in the tube is uniformly
distributed over its circular cross-section of radius  $a$.
 Other possible current  distributions just slightly modify the constant  $5/4$
(see \cite{TitDem99b}). 
   For the assumed small radius of the flux tube ($a \ll R$) the
toroidal field $B_{\theta}$ is approximately homogeneous in the vicinity of the tube.
   This enables us to relate the radius  $a$ to the total number  $n_{\rm
t}$ of field line turns in the tube as follows:
 \begin{eqnarray}
  n_{\rm t} \approx {\mu_{0} I R \over 2 \pi a^{2} |B_{\theta}|} = {I
R^{2} \over |I_{0}| a^{2}}  \, .  
	\label{Nt}
 \end{eqnarray}
 In the above mentioned scenario of the emerging flux tube only the
equilibrium current  $I$ and minor radius $a$ change with growing major
radius  $R$ according to (\ref{Ieq}) and (\ref{Nt}), while  $n_{\rm
t}$ and the other parameters of the model  $q$,  $d$, $L$,  and
$I_{0}$ remain constant.
 At each moment of such an evolution Eqs.~(\ref{Hp}), (\ref{Psp}), (\ref{Psq}),  and
(\ref{PsI})--(\ref{Nt}) describe the Hamiltonian dynamic system
(\ref{Heq1})--(\ref{Heq2}) corresponding to the magnetic field outside
the flux tube. 

\begin{figure}[!ht] 
   \begin{center}
    \leavevmode
  \centerline{\epsfig{file=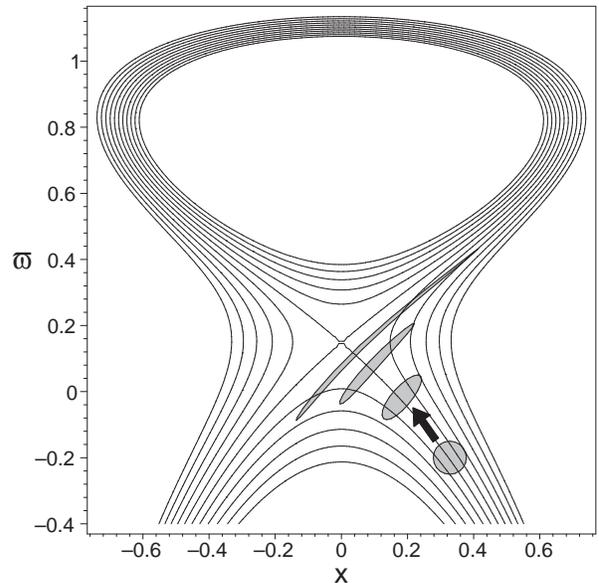,width=7.5cm}}
  \end{center}
   \caption{\em  The phase volume at four different equidistant
moments: the initial volume is represented by a small circular, which can
be considered as the corresponding photospheric cross-section of the
flux tube starting near the center of the configuration;  the final
highly stretched volume corresponds to the moment
when the field line, passing through the X-point of the poloidal field,
reaches the photosphere.}
    \label{f:phfl}
 \end{figure}

  The Hamiltonian in this region  has a hyperbolic critical point,
which is in fact the X-point of the poloidal magnetic field.
  Figure \ref{f:phfl} shows the corresponding phase portrait of the
system for the following set of
parameters:  $n_{\rm t} = 5$,   $q=100 \: {\rm T\cdot Mm^{2}}$,
$L=50 \: {\rm Mm}$,  $R=110 \: {\rm Mm}$,  $I_{0}= -7
\:  {\rm TA}$ (cf. \cite{TitDem99b}b).
 This portrait is a result of projection $(x,\varpi, \theta)
\rightarrow  (x,\varpi)$ of several magnetic field lines.
  It can also be considered as a set of contours  $H={\rm const}$,
since our autonomous phase flow conserves the Hamiltonian.
   Taking into account Eq.  (\ref{Hp}) or more general
Eq. (\ref{Ham}), one can obtain another useful interpretation of the
portrait, namely,  as a cross-section $\theta={\rm const}$ of the corresponding
poloidal magnetic  surfaces $\Psi={\rm const}$ shown in
the coordinates $(x,\varpi)$.
  These coordinates differ from Cartesian ones only by the logarithmic
transformation  ($\varpi=\ln r_{\perp}$) of the vertical axis, so
Figure \ref{f:phfl} yields a rather realistic information about the
shape of these surfaces.

  The important property of autonomous Hamiltonian systems is the
conservation of the corresponding phase volume in the phase flow.
  In our case this volume is two-dimensional, so the area of the plane
$(x,\varpi)$ is conserved in such a flow.
 This property in combination with the presence of the hyperbolic
critical point, provides a mechanism of the QSL formation
in our configuration.
  Indeed, any cross-section of an elementary flux tube corresponds to 
an area in the phase plane and the variation of this cross-section
along such a tube corresponds to an evolution of the appropriate phase
area due to the phase flow.
   Given a small circle near one of the separatrices on the
phase portrait as an initial area,  its area-conserving evolution
with the divergent character of the phase flow
near the critical point will inevitably cause a stretching and compressing 
of this area across and along, respectively, the direction of the flow
(see Figure \ref{f:phfl}).
  This means that the elementary flux tubes starting nearby the poloidal
separatrix surfaces have to experience especially strong flattening,
  which  is a characteristic feature of the field lines in QSL.

  One can also expect from the above consideration that the maximal
value of  $Q \equiv Q_{\rm X}$ is achieved at the field line passing through the
X-point of poloidal magnetic field.
 It is possible to derive an analytical expression for  $Q_{\rm X}$ 
by first linearizing Eqs. (\ref{Heq1})--(\ref{Heq2})  at the hyperbolic
critical point of  $H$ and then solving them in order to calculate the
corresponding derivatives for  $Q_{X}$.
  The final result reads
 \begin{eqnarray}
 Q_{\rm X} = 2 +  \left(  4 + {d^{4} L^{-2} r^{-2}_{\rm X}  \over 
L^{2} r^{2}_{\rm X}  - d^{2}}  
 \right) \, \sinh^{2}   ( \lambda \, \Delta\theta )\, ,
	\label{QX}
 \end{eqnarray}
where  $r_{\rm X}$ ( $< R/L$) stands for the dimensionless radius  $r_{\perp}$
corresponding to the X-point of the poloidal field.
 The value  $\lambda$ is given by
 \begin{eqnarray}
  \lambda =  r_{\rm X} \left.  {\partial^{2} H  \over \partial
x^{2}} \right|_{x=0, \varpi=\ln r_{\rm X}} 
	\label{lmbd}  
 \end{eqnarray} 
 and according to (\ref{Bax}), (\ref{Btp}) and (\ref{Hp}) it
characterizes the strength of the poloidal field on the distances
$\sim L r_{\rm X}$ in comparison with the toroidal field at the X-point.
  The value
 \begin{eqnarray}
  \Delta\theta  =  \pi -2 \arcsin  \left( {d \over L  r_{\rm X} }
\right) \, 
	\label{Dth}
 \end{eqnarray}
determines the arc length of the coronal part of the field line
passing through the X-point.

 \begin{figure}[!ht] 
   \begin{center}
    \leavevmode
  \centerline{\epsfig{file=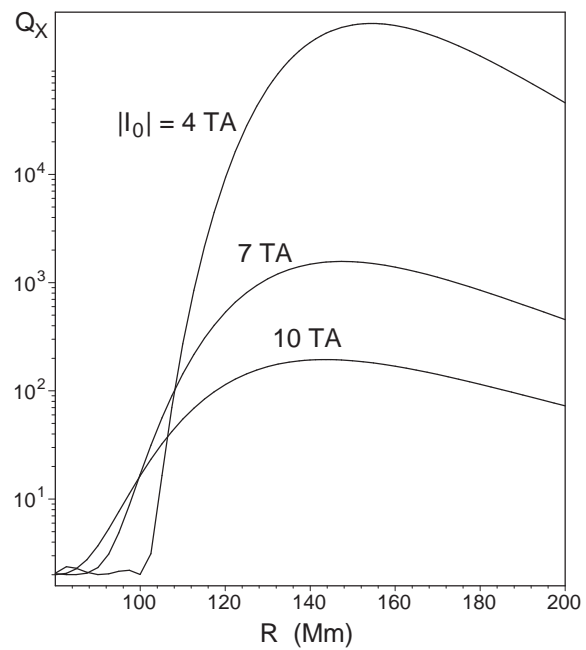,width=7.5cm}}
  \end{center}
   \caption{\em  The dependence of $Q\equiv Q_{\rm X}$ on the radius
of the flux tube $R$ for the field line passing through the X-point
of poloidal magnetic field at different values of the line current
$I_{0}$ (Figure \ref{schm}).}
    \label{f:QX}
 \end{figure}

  Eq.~(\ref{QX}) shows that  $Q_{\rm X}$ grows exponentially with
$\lambda \,\Delta \theta$, so  $Q_{\rm X}$ strongly depends on the
relationship between poloidal and toroidal components in the vicinity of
the X-point and the length of the field line passing through this
point.
  This value has been computed for the above mentioned
parameters and  $d=50 \: {\rm Mm}$ in the scenario of the emerging flux
tube.
 The dependence  $Q_{\rm X}(R)$ at several different values of
$I_{0}$ (see Figure (\ref{f:QX})) suggests  that the QSL has to appear 
in the twisted configuration when the major radius of the
flux tube reaches $R \ga 100 \: {\rm Mm}$.
  This corresponds approximately to the moment when the genuine
separatrix surfaces (caused by BPs) disappear (\cite{TitDem99b}b).

 \section{CONCLUSIONS}

  In this paper, new results concerning the
theory of the quasi-separatrix layers (QSLs) in coronal magnetic
fields are reported.
  We have revised the previous definition of QSLs  (\cite{PriDem95})
by analyzing the local geometrical properties of the field line
connectivity.
  This analysis provides a new geometrical measure 
$Q$, which determines the degree of flattening of the elementary flux
tubes.
  Contrary to the so-called norm  $N$ proposed previously by
\cite*{PriDem95}, the new measure  $Q$ yields the same results for both
positive and negative magnetic polarities on the photosphere.
  In other words,  $Q$ characterizes the magnetic connectivity itself
rather than the corresponding field line mappings (from positive to
negative polarity or vice versa).

  The old and new measures are related by a very simple formula
$Q=N^{2} / \Delta$, in which  $\Delta$ is the Jacobian of the
corresponding field line mapping.
  Thus, if  $\Delta$ varies on the photosphere as strong as  $N$,  the 
new and old measure must yield different locations for QSLs.

  We have applied this theory to the model of twisted magnetic
configuration proposed in (\cite{TitDem99a}b) for the analysis of
magnetic topology in sigmoid-like solar flares.
  It has been demonstrated that the appearance of QSLs in such a model 
is related with the Hamiltonian structure of the modeling
magnetic field: the conservation of the phase area in the
corresponding phase flow and the presence of the X-point in the
poloidal magnetic field provides the mechanism for the formation of
QSLs in this configuration.
  The results of computations of  $Q$ suggest that the QSL must be
formed in such a configuration approximately after that the
genuine separatrix surfaces disappear when the twisted flux tube
emerges high enough into the corona.

 \section*{ACKNOWLEDGMENTS}
  V.S.~Titov and G.~Hornig gratefully acknowledge financial support from
Volkswagen-Foundation.

\end{document}